\RequirePackage{fix-cm}
\documentclass[smallextended]{svjour3}       
\smartqed  
\usepackage{graphicx}
\usepackage{amsmath}
\usepackage{amssymb}
\usepackage{url}
\usepackage{bm}
\usepackage{mathrsfs}

\begin{document}

\title{Study on upper limit of sample size for a two-level test in
  NIST SP800-22
\thanks{
  The author is supported by Grant-in-aid for Science Research, 
  Nos. 16K13750, 17K14234, and 18K03213.
}
}


\author{Hiroshi Haramoto
}


\institute{H. Haramoto \at
  Faculty of Education, Ehime University, 3 Bunkyocho, Matsuyama, Ehime 790-8577, Japan \\
  \email{haramoto@ehime-u.ac.jp}           
}

\date{Received: date / Accepted: date}

\maketitle

\begin{abstract}
NIST SP800-22 is one of the most widely used statistical testing tools 
for pseudorandom number generators (PRNGs). 
This tool consists of 15 tests (one-level tests) and two additional tests
(two-level tests). 
Each one-level test provides one or more $p$-values.
The two-level tests measure
the uniformity of the obtained $p$-values
for a fixed one-level test.
One of the two-level tests categorizes the $p$-values into ten intervals
of equal length, and apply a chi-squared goodness-of-fit test.
This two-level test is often more powerful than one-level tests,
but sometimes it rejects even good PRNGs when the sample size at
the second level is too large,
since it
detects approximation errors in the computation of $p$-values.

In this paper, we propose 
a practical upper limit of the sample size in this two-level test,
for each of six tests appeared in SP800-22.
These upper limits are derived by the chi-squared discrepancy 
between the distribution of the approximated $p$-values and 
the uniform distribution $U(0, 1)$. 
We also computed a ``risky'' sample size at the second level
for each one-level test. Our experiments show that
the two-level test with the proposed 
upper limit gives appropriate results, while
using the risky size often rejects even good PRNGs.

We also propose another improvement:
to use the exact probability for the ten categories
in the computation of goodness-of-fit at the two-level test.
This allows us to increase the sample size at the second level,
and would make the test more sensitive than
the NIST's recommending usage.
\end{abstract}

\section{Introduction}
\label{intro}

Statistical testing is a common way to evaluate whether 
the outputs of a pseudorandom number generator (PRNG) imitate 
independent random variables from the uniform distribution 
over the interval $[0,1]$, or over the integers in an interval 
(e.g. $\{0,1\}$).

There are many statistical tests for PRNGs, for example, 
TestU01 by L'Ecuyer and Simard is the most comprehensive test suite
\cite{L'Ecuyer:2007:TCL:1268776.1268777}, 
and recently PractRand \cite{practrand} has been published. 
In this paper, we investigate a test suite, named SP800-22,
that is proposed by National Institute of Standards and Technology (NIST) 
\cite{Bassham:2010:SRS:2206233} for pseudorandom bit generators (PRBGs). 
NIST SP800-22 is a standard test suite, especially in cryptography.  
This suite consists of 15 tests called one-level tests and 
two additional tests called two-level tests.
One of the two-level tests categorizes the $p$-values
into ten equal-length subintervals of $[0,1]$, 
and apply a chi-squared goodness-of-fit (GOF) test. 
This two-level test permits us to apply the test with a larger total 
sample size, and hence, it is often more powerful than one-level tests. 
On the other hand, the two-level test tends to reject even good PRNGs
because of approximation errors in computation of the $p$-values at the 
first level.
Hence giving upper limit of sample size at the second level makes 
the two-level test more reliable: 
we can maximize the power and avoid erroneous rejections of 
the two-level test. 
Further descriptions and explanations of two-level tests can be found in 
\cite{10.1007/978-3-642-04107-5_26,Knuth:1997:ACP:270146,rLEC92a,MR1310607,doi:10.1080/00949659708811859,4252919,6135498,Yamaguchi}.

However, NIST recommends that the sample size at the second level 
should be on the order of the inverse of the significance level, 
without any mathematical justifications. 
Pareschi et al. \cite{6135498} give the upper limit of the two-level 
test for the Frequency test and the Runs test in NIST SP800-22. 
Their research is based on Berry-Ess\'{e}en inequality 
\cite{Shiryaev:1995:PRO:217651}, 
but it seems difficult to derive upper limits of the other one-level tests. 
By contrast, we propose a practical upper limit of the 
sample size at the second level for each of the Frequency test, 
the Binary Matrix Rank test and the Runs test 
\cite{10.1007/978-3-319-91436-7_15}. 
Those upper limits are derived by 
the chi-squared discrepancy between the distribution of 
approximated $p$-values and the uniform distribution. 
For example, we show that 
the upper limit of sample size at the second level test for 
the Frequency test 
is approximately $125{,}000$ when the sample size at the first level 
is $10^6$.  
This value is larger than that recommended by NIST, which is $10^3$--$10^4$.  
We also show that the upper limits increase the power of the two-level test. 

The aim of this paper is to extend the previous work \cite{10.1007/978-3-319-91436-7_15},   
in particular, we propose a practical upper limit of 
the sample size at the second level for each of the following six tests 
in NIST SP800-22: the Longest-Run-of-Ones in a Block, 
the Overlapping Template Matching test, 
the Linear Complexity test, the Random Excursions test, 
the Frequency test within a Block, 
and the Discrete Fourier Transform test. 
The results indicate that appropriate upper limits heavily depend on 
the one-level tests and sample sizes at the first level.

The rest of this paper is organized as follows. 
In Section 2, we review the test suite NIST SP800-22 and 
the chi-squared discrepancy, and explain how to determine
the upper limits of the sample size at the second level. 
In Section 3, we present upper limits of several tests and 
experimental results. 
In Section 4, we give an approximation of the distribution of 
$p$-values of one-level tests using a Monte Carlo simulation. 
In Section 5, we propose another improvement using
the exact probability for the ten categories in the computation 
of the GOF in the two-level test. 

\section{Two-level test in NIST SP800-22 and chi-squared discrepancy}
\label{sec:2}
In this section, we briefly explain NIST SP800-22 and the chi-squared discrepancy. 

NIST SP800-22 consists of 15 statistical tests called one-level tests. 
Each one-level test looks for empirical evidence against a null hypothesis 
$\mathcal{H}_{0,first}$ that random variables $B_1$, $B_2$ $\ldots$, $B_n$ are i.i.d. 
over the two-element set $\{0, 1\}$, i.e., 
$$
\mathcal{H}_{0,first} : B_1, \ldots, B_n \sim_{i.i.d.} Binom(1, 1/2). 
$$

Let $T$ be a statistic of a statistical test. 
When testing the randomness, we assume that the distribution of $T$ 
under $\mathcal{H}_{0,first}$ is known or well-approximated by a computable formula. 
We denote by $F$ the cumulative distribution function of $T$.
For a realization $(b_1, \ldots, b_n) \in \{0,1\}^n$, where $b_i$ is the 
$i$-th output of the tested PRBG, the test rejects $\mathcal{H}_{0,first}$ 
and the PRBG if the probability (called the $p$-value) 
$$
\mathbb{P}(T(b_1, \ldots, b_n) \leq T(B_1, \ldots, B_n)) 
= 1-F(T(b_1, \ldots, b_n)) 
$$
is much too close to either 0 or 1; we call $n$ the first sample size. 

If the $p$-value is very small (e.g. $<10^{-10}$), then it is clear that
the PRBG fails the test. However, if the $p$-value is suspicious but 
does not clearly indicate rejection (e.g. $\approx 10^{-4}$), 
it is difficult to judge. 

In order to avoid such difficulties, a two-level test is often used. 
We fix a one-level test. At the first level, we apply the one-level test $N$ times 
to disjoint parts of a sequence generated by the PRBG, yielding $N$ $p$-values. 
At the second level, we compare the empirical distribution of those $p$-values 
to the expected distribution via a GOF test; 
we call $N$ the second sample size.  

NIST SP800-22 includes two two-level tests. 
In this paper we investigate the test referred to as the uniformity test, 
which is detailed below.
Let $\nu$ be a positive integer. 
We denote by $I_0$, $\ldots$, $I_{\nu}$ the subintervals of $[0,1]$ defined by 
the following:
\begin{align*}
I_i &:= [i/(\nu+1), (i+1)/(\nu+1)) ~~ (i=0,1,\ldots,\nu-1), \mbox{and}\\
I_{\nu} &:= [\nu/(\nu+1), 1]. 
\end{align*}
Let $Y_i$ be the number of $p$-values that fall in the subinterval 
$I_i$, $i=0, \ldots, \nu$. 
If the random vector $(Y_0, \ldots, Y_{\nu})$ conforms to 
$Multi(N; p_0, \ldots, p_{\nu})$,  
the multinomial distribution with $N$ trials and probability 
$(p_0, \ldots, p_{\nu})$, then the chi-squared statistic 
$$
\chi^2 := \sum_{i=0}^{\nu} \frac{(Y_i - N p_i)^2}{Np_i} 
$$
approximately conforms to the chi-squared distribution with $\nu$ degrees of freedom 
for large $N$. 

It is often assumed that the distribution of the $p$-values
of a one-level test is
uniform over $[0, 1]$. 
(Note that this is incorrect if $T$ comes from a discrete
probability distribution.)
Under this uniformity assumption, the two-level test examines a null hypothesis 
$$
\mathcal{H}_{0,second} : p_0 = 1/(\nu+1), \ldots, p_{\nu} = 1/(\nu+1).
$$
For a realization $\chi^2_{obs}$ of $\chi^2$, NIST SP800-22 defines the $p$-value 
at the second level by $\mathbb{P}(\chi^2_{obs} \leq X)$, 
where $X$ is a random variable that conforms to the chi-squared distribution with 
$\nu$ degrees of freedom. If the $p$-value at the second level is 
less than a pre-specified significance level, 
then we reject $\mathcal{H}_{0, second}$ and the PRBG. 

Two-level tests are said to be more sensitive than one-level tests. 
However, such tests may lead to erroneous rejection, which is explained as follows:
all the 15 one-level tests in NIST SP800-22 approximately 
compute $p$-values using continuous distributions 
instead of the actual discrete distributions. 
For example, the test statistic of the Frequency test in NIST SP800-22,
conforms to a binomial distribution, but the test computes 
approximated $p$-values using a normal distribution.
As a result, the probability $q_i$ that the approximated $p$-values 
fall in the interval $I_i$ differs from $1/(\nu+1)$. 
The test statistic $\chi^2$ has a larger deviation if $N$ is larger, 
and the null hypothesis $\mathcal{H}_{0,second}$ is more likely to be rejected.

In order to decide an upper limit of second sample size, 
we here quantify the discrepancy between the distribution of approximated $p$-values of a one-level test $\{q_i\}$ and the null distribution $\{p_i\}$ under $\mathcal{H}_{0,second}$. 
Assume that $(Y_0, \ldots, Y_{\nu})$ actually conforms to $Multi(N; q_0, \ldots, q_{\nu})$,
but tested by a chi-squared GOF test, assuming the null hypothesis that 
$(Y_0, \ldots, Y_{\nu})$ conforms to $Multi(N; p_0, \ldots, p_{\nu})$.
In this case, Matsumoto and Nishimura \cite{MR1958868} showed the following 
inequality:
\begin{align}
  \label{eq:bound}
  |E(\chi^2)-(\nu+N \delta)| \leq 
  \nu \max_{i=0, 1, \ldots , \nu} \left| 1-\frac{q_i}{p_i} \right|. 
\end{align}
Here a distance-like function 
$$
\displaystyle \delta := \sum_{i=0}^{\nu} \frac{(q_i-p_i)^2}{p_i}
$$
refers to the chi-squared discrepancy of $\{q_i\}$ from $\{p_i\}$.
Note that the chi-squared statistic $\chi^2$ is
  known to conform to the noncentral chi-squared distribution
  having $\nu$ degrees of freedom with noncentrality parameter $N\delta$
  \cite{tiku}.

Inequality (\ref{eq:bound}) implies that
$\chi^2$ value is shifted by $N\delta$ in average. 
In addition, the expectation of $\chi^2$ 
corresponds to the $p$-value $\alpha \in (0, 1)$ when
$$
N \approx \frac{\chi^2_{\nu}(\alpha)-\nu}{\delta},
$$
where $\chi^2_{\nu}(\alpha)$ is the upper $100 \alpha$-th percentile
of the chi-squared distribution with $\nu$ degrees of freedom for $\alpha$.

In this paper, we only deal with the case $\nu=9$ and the significance level $0.0001$, 
which are initial values in NIST SP800-22.
Let
$$
u=\max_{i=0, \ldots , \nu}   \left| 1-\frac{q_i}{p_i} \right|. 
$$
The two-level test with the second sample size 
$$
N_{0.0001}:= \left\lceil \frac{\chi^2_{\nu}(0.0001)-\nu+u\nu}{\delta} \right\rceil  
$$
tends to reject PRBGs even if the tested PRBGs are ideal:
at the sample size $N_{0.0001}$, the expectation of the $\chi^2$-value
corresponds to the $p$-value
$$
\mathbb{P}(N_{0.0001}\delta + \nu - u\nu \leq X) \approx 0.0001,
$$
where $X$ is a random variable conforming to the chi-squared distribution
with $\nu$ degrees of freedom. On the other hand, if we take
$$
N_{0.25}:= \left\lfloor \frac{\chi^2_{\nu}(0.25)-\nu-u\nu}{\delta} \right\rfloor
$$
as the second sample size, such erroneous rejections are unlikely to occur:
the expectation of the $\chi^2$-value corresponds to the $p$-value
$$
\mathbb{P}(N_{0.25}\delta + \nu + u\nu \leq X) \approx 0.25.
$$


For these reasons, $N_{0.0001}$ and 
$N_{0.25}$ are called the risky sample size and the
safe sample size at the second level, respectively, in \cite{MR1958868}.
The name ``safe sample size'' is somehow misleading, since
the Inequality (1) implies that the 
distribution of $\chi^2$ is considerably different from 
a chi-squared distribution even if one takes $N_{0.25}$
as the second sample size, and hence 
the probability of an erroneous rejection 
for an ideal random number generator is higher 
than a pre-specified significance level.
Indeed, a numerical computation based 
on the noncentral chi-squared distribution shows that
if the significance level is set to be $0.0001$ 
(the default value for NIST tests) with 
degree of freedom $9$,
the probability of rejection due to too high 
$\chi^2$-value is about $0.001205194$, which 
is much higher than $0.0001$. However, 
if the NIST test procedure is used as it is,
the probability of the rejection is around $0.0012$,
which would not result in the rejection of an ideal generator.
The name ``safe'' means only in this context: the 
safe sample size is a thumb nail for the upper limit
of the second sample size for the NIST test procedure.

\section{Computing the distributions of $p$-values of some statistical tests}
\paragraph{Three tests in SP800-22 based on chi-squared test.}
First, we consider the two-level test for 
the following three one-level tests:
the test for the Longest-Run-of-Ones in a Block, the Overlapping Template Matching test, 
and the Linear Complexity test. Note that we apply a modification to 
the test for the Longest-Run-of-Ones in a Block
to improve the approximation of $p$-values \cite{HARAMOTO201966}.

Below, we explain the common structure of these one-level tests.
Each one-level test divides the $n$-bit sequence into 
$n_b=\lfloor n/m \rfloor$ blocks $P_1$, $\ldots$, $P_{n_b}$ of $m$ bits.
According to a certain property which the test investigates,  
the test classifies $P_i$'s into $k+1$ classes $C_0, \ldots, C_k$.
Let $X_0,\ldots, X_k$ be the numbers of 
the blocks classified in $C_0,\ldots,C_k$, 
respectively.
The test compares an observed frequency $\bm{X}=(X_0, \ldots, X_k)$
to the theoretical one: NIST describes that the theoretical distribution of 
$\bm{X}$ is $Multi(n_b; \pi_0, \ldots, \pi_k)$ explicitly 
\cite{Bassham:2010:SRS:2206233}.
The test statistic of the one-level test is 
\begin{align}
\label{eq:teststatistic}
T(\bm{X})=T(X_0, \ldots, X_k) := 
\sum_{i=0}^k \frac{(X_i-n_b\pi_i)^2}{n_b\pi_i}.
\end{align}

Let us denote by $S$ the set of the realizations of $\bm{X}$, i.e., 
\begin{align}
\label{eq:samplespace}
S = \left\{ \bm{x} = (x_0, \ldots, x_k) \in \mathbb{Z}_{\geq 0}^{k+1}
\mid x_0+\cdots+x_k = n_b \right\}.
\end{align}
For a realization $\bm{x} \in S$, the one-level test approximates its $p$-value
by $\mathbb{P}(T(\bm{x}) \leq X)$, where $X$ is a random variable that follows 
the chi-squared distribution with $k$ degrees of freedom. 
Therefore, the probability $q_i$ that the approximated $p$-values fall in the 
subinterval $I_i$ 
is 
$$
q_i = \sum_{\bm{x} \in S, ~~ \mathbb{P}(T(\bm{x}) \leq X) \in I_i} 
\dfrac{n_b!}{x_0!\cdots x_k!} \pi_0^{x_0} \cdots \pi_k^{x_k},  
i=0, 1, \ldots, \nu.  
$$

Exhaustive computation derives the values of $q_i$'s.
Table 1 shows the values of $q_i$'s, $\delta$, $N_{0.25}$, and $N_{0.0001}$
when the first sample size is $n=10^6$.
The block size $m$ is indicated in the second row of Table 1, 
and the other test parameters are the initial values in NIST SP800-22.

\begin{table}[htpb]
  \begin{center}
    \caption{Values of $q_i$'s, $\delta$, $N_{0.25}$ and 
      $N_{0.0001}$ of three one-level tests}
    \begin{tabular}{cccc}
      \hline
      Test & Longest & Overlap & Linear \\
      $m$ & $10{,}000$ & $1{,}032$ & $5{,}000$ \\ \hline
      $q_0$ & 0.0984739 & 0.0998142 & 0.0992755 \\
      $q_1$ & 0.0993067 & 0.0999758 & 0.0958139 \\
      $q_2$ & 0.1003668 & 0.1000190 & 0.0965409 \\
      $q_3$ & 0.1008263 & 0.1000541 & 0.0994559 \\
      $q_4$ & 0.1011301 & 0.1000879 & 0.1029601 \\
      $q_5$ & 0.1010720 & 0.1000979 & 0.1013597 \\
      $q_6$ & 0.1007868 & 0.1000902 & 0.1033444 \\
      $q_7$ & 0.1004239 & 0.1000561 & 0.1025969 \\
      $q_8$ & 0.0994782 & 0.0999548 & 0.1004993 \\
      $q_9$ & 0.0981354 & 0.0998500 & 0.0981534 \\ \hline
      $\delta$ & $1.060097\times10^{-4}$ & $9.150630\times10^{-7}$ & $6.250910\times10^{-4}$ \\
      $N_{0.25}$ & $20{,}950$ & $2{,}592{,}207$ & $3{,}218$ \\
      $N_{0.0001}$ & $234{,}769$ & $27{,}032{,}746$ & $40{,}149$ \\ \hline
    \end{tabular}    
  \end{center}
\end{table}

In order to justify the name of the risky/safe sample sizes, 
we apply the two-level test to Mersenne Twister (MT) \cite{DBLP:journals/tomacs/MatsumotoN98} 
and a PRNG from the SHA-1 algorithm (SHA-1) with five different initial random seeds. 
We take the second sample sizes approximately $N_{0.25}$ and $N_{0.0001}$. 
We assume that both MT and SHA-1 are good generators, and  
thus the empirical distribution of $\chi^2$ are expected to conform to 
the theoretical distribution described in Section 2. 
Because the Linear Complexity test always yields very small $p$-values
when tested PRNGs have $\mathbb{F}_2$-linearity, 
we apply the two-level test for the Linear Complexity test only to SHA-1.

Table 2 shows the resulting $p$-values of the two-level test. 
The table shows that $N_{0.0001}$'s yield small $p$-values, 
and then some rejections occur. On the other hand, 
$N_{0.25}$'s give no rejections at the second level. 

\begin{table}[htpb]
  \begin{center}
    \caption{$p$-values of the two-level tests on MT and SHA-1}
    \begin{tabular}{ccccccccc}
      \hline 
      Test & PRNG & $N$ & 1st & 2nd & 3rd & 4th & 5th \\ \hline
      Longest & MT & 21{,}000 & 
      3.75e-01 & 5.76e-01 & 7.02e-02 & 9.76e-03 & 6.39e-01  \\ 
      & & 235{,}000 &
      7.82e-03 & 6.93e-08 & 1.74e-08 & 1.05e-03 & 7.24e-07 \\ \cline{2-8}
      & SHA-1 & 21{,}000 & 
      4.82e-01 & 7.64e-01 & 9.77e-01 & 5.48e-04 & 9.64e-01  \\ 
      & & 235{,}000 &
      3.89e-04 & 9.18e-05 & 2.11e-02 & 2.41e-07 & 6.02e-05 \\ \hline
      \hline 
      Overlap & MT & 2{,}600{,}000 & 
      1.47e-01 & 9.69e-01 & 1.41e-01 & 7.27e-02 & 4.01e-02  \\ 
      & & 27{,}033{,}000 &
      1.17e-02 & 2.71e-03 & 1.12e-02 & 3.20e-03 & 8.53e-05 \\ \cline{2-8}
      & SHA-1 & 2{,}600{,}000& 
      2.00e-01 & 8.68e-02 & 4.36e-01 & 2.68e-02 & 6.02e-01  \\ 
      & & 27{,}033{,}000 &
      1.26e-05 & 1.63e-05 & 1.14e-04 & 1.19e-03 & 1.08e-03 \\ \hline
      \hline 
      Linear & SHA-1 & 3{,}200 & 
      4.94e-01 & 9.18e-01 & 1.05e-01 & 1.96e-01 & 6.77e-02  \\ 
      & & 40{,}200 &
      1.96e-03 & 1.40e-10 & 1.57e-05 & 7.38e-06 & 1.00e-04 \\ \hline
    \end{tabular}
  \end{center}
\end{table}

\paragraph{The Random Excursions test.}
We consider the Random Excursions test. 
This test yields simultaneously eight $p$-values (one for each test parameter 
$x = \pm 1$, $\pm 2$, $\pm 3$ and $\pm 4$) 
when the number of cycles $J$, which is determined by the tested sequence, 
is greater than or equal to $500$, and yields no results when $J<500$.

For the sake of simplicity, in this paper, we alter the test procedure slightly:
when $J=500$, the test yields eight $p$-values and discards the remaining bits,
and when $J<500$, the test yields no results.

Table 3 includes the results of exhaustive computation of $q_i$'s, etc.
Tables 4 and 5 show the resulting $p$-values of this two-level test
on MT and SHA-1, respectively.

\begin{table}[htpb]
  \begin{center}
    \caption{The values of $q_i$'s, $\delta$, $N_{0.25}$ and $N_{0.0001}$
    of the Random Excursions test}
    \begin{tabular}{ccccc}
      \hline
      $x$ & $\pm 1$ &  $\pm 2$ & $\pm 3$ & $\pm 4$ \\ \hline
      $q_0$ & 0.0994313 & 0.0993668 & 0.0989417 & 0.0986037 \\
      $q_1$ & 0.0992651 & 0.0989767 & 0.0976982 & 0.0959258 \\
      $q_2$ & 0.0999442 & 0.0998576 & 0.0994305 & 0.0984995 \\
      $q_3$ & 0.1006237 & 0.1000796 & 0.1005629 & 0.1005897 \\
      $q_4$ & 0.0999540 & 0.1006024 & 0.1005381 & 0.1014219 \\
      $q_5$ & 0.1006994 & 0.1006013 & 0.1014175 & 0.1016747 \\
      $q_6$ & 0.1007053 & 0.1006613 & 0.1010161 & 0.1026947 \\
      $q_7$ & 0.1001240 & 0.1004919 & 0.1014867 & 0.1011569 \\
      $q_8$ & 0.0998671 & 0.0998926 & 0.0994729 & 0.1013548 \\
      $q_9$ & 0.0993858 & 0.0994698 & 0.0994353 & 0.0980782 \\ \hline
      $\delta$ & $2.654570\times10^{-5}$ & $3.171128\times 10^{-5}$ & $1.319765\times 10^{-4}$
      & $4.010343\times10^{-4}$\\  
      $N_{0.25}$ & $87{,}494$ & $72{,}423$ & $16{,}530$ & $5{,}042$ \\
      $N_{0.0001}$ & $933{,}714$ & $782{,}436$ & $188{,}876$ & $62{,}555$ \\ \hline
    \end{tabular}    
  \end{center}
\end{table}

\begin{table}[htpb]
  \begin{center}
    \caption{$p$-values of the two-level test for 
      the Random Excursions test on MT}
    \begin{tabular}{ccccccc}
      \hline 
      $x$ & $N$ & 1st & 2nd & 3rd & 4th & 5th \\ \hline
      $-4$ & 5{,}000 
      & 1.85e-01 & 7.85e-01 & 1.12e-01 & 1.54e-01 & 3.28e-01 \\
      & 63{,}000 
      & 3.14e-05 & 1.33e-04 & 9.75e-05 & 1.39e-06 & 2.01e-02 \\ \hline
      $-3$ & 16{,}500  
      & 2.03e-01 & 9.19e-01 & 1.36e-01 & 2.38e-01 & 7.51e-01 \\
      & 190{,}000 
& 4.00e-03 & 5.79e-03 & 5.64e-05 & 1.52e-05 & 1.11e-05 \\ \hline
      $-2$ & 72{,}000
  & 4.18e-01 & 1.98e-01 & 2.05e-01 & 2.65e-02 & 5.68e-02 \\
      & 783{,}000 
& 9.96e-05 & 1.96e-05 & 7.50e-03 & 3.38e-08 & 6.24e-04 \\ \hline
      $-1$ & 87{,}000  
& 2.85e-01 & 5.06e-01 & 5.59e-01 & 1.99e-01 & 7.06e-02 \\
      & 934{,}000 
& 5.22e-10 & 3.91e-05 & 2.59e-07 & 1.07e-08 & 2.63e-09 \\ \hline
      $1$ & 87{,}000  
& 5.79e-01 & 5.98e-01 & 5.57e-01 & 2.98e-01 & 7.55e-01 \\
      & 934{,}000 
& 3.31e-07 & 9.59e-06 & 2.88e-04 & 4.01e-07 & 2.26e-07 \\ \hline
      $2$ & 72{,}000  
& 2.07e-02 & 4.35e-01 & 3.94e-01 & 9.56e-01 & 3.39e-01 \\
      & 783{,}000 
& 3.91e-06 & 2.78e-09 & 9.24e-04 & 1.01e-03 & 3.92e-05 \\ \hline
      $3$ & 16{,}500  
& 1.34e-01 & 2.68e-01 & 5.41e-01 & 1.02e-01 & 3.56e-02 \\
      & 190{,}000 
& 6.92e-05 & 3.82e-08 & 3.10e-07 & 7.69e-05 & 7.76e-09 \\ \hline
      $4$ & 5{,}000  
& 4.54e-01 & 4.99e-02 & 4.44e-01 & 1.64e-01 & 8.52e-01 \\
      & 63{,}000 
& 3.70e-04 & 5.06e-03 & 1.05e-03 & 1.33e-04 & 6.69e-05 \\ \hline
    \end{tabular}    
  \end{center}
\end{table}

\begin{table}[htpb]
  \begin{center}
    \caption{$p$-values of the two-level test for
      the Random Excursions test on SHA-1}
    \begin{tabular}{ccccccc}
      \hline 
      $x$ & $N$ & 1st & 2nd & 3rd & 4th & 5th \\ \hline
      $-4$ & 5{,}000 & 
      7.88e-01 & 2.16e-01 & 1.50e-02 & 1.28e-01 & 9.20e-03  \\ 
      & 63{,}000 &
      6.38e-06 & 4.90e-10 & 1.08e-02 & 4.41e-03 & 4.92e-07 \\ \hline
      $-3$ & 16{,}500 & 
      6.10e-02 & 6.61e-02 & 3.81e-01 & 8.15e-01 & 4.11e-01  \\ 
      & 190{,}000 &
      4.65e-07 & 2.50e-07 & 6.18e-03 & 5.47e-05 & 4.77e-05 \\ \hline
      $-2$ & 72{,}000 & 
      6.34e-01 & 5.55e-03 & 7.88e-01 & 3.89e-01 & 5.20e-01  \\ 
      & 783{,}000 &
      3.36e-06 & 3.81e-11 & 5.56e-05 & 9.00e-08 & 8.73e-04 \\ \hline
      $-1$ & 87{,}000 & 
      4.44e-01 & 1.38e-03 & 7.31e-01 & 5.50e-01 & 6.14e-01  \\ 
      & 934{,}000 &
      5.48e-11 & 8.58e-10 & 2.34e-08 & 1.72e-05 & 1.13e-03 \\ \hline
      $1$ & 87{,}000 & 
      5.67e-02 & 7.76e-01 & 6.82e-02 & 2.09e-01 & 2.93e-01  \\ 
      & 934{,}000 &
      9.77e-12 & 1.50e-09 & 5.04e-09 & 1.18e-04 & 3.65e-15 \\ \hline
      $2$ & 72{,}000 & 
      6.83e-04 & 1.77e-01 & 1.05e-02 & 2.34e-01 & 1.99e-01  \\ 
      & 783{,}000 &
      6.88e-11 & 5.54e-07 & 1.20e-04 & 8.55e-10 & 7.38e-05 \\ \hline
      $3$ & 16{,}500 & 
      8.06e-01 & 7.94e-01 & 1.12e-01 & 5.93e-01 & 2.15e-01  \\ 
      & 190{,}000 &
      1.77e-04 & 2.52e-03 & 5.37e-05 & 5.65e-08 & 3.11e-04 \\ \hline
      $4$ & 5{,}000 & 
      3.35e-01 & 2.97e-01 & 8.31e-02 & 2.69e-02 & 5.77e-01  \\ 
      & 63{,}000 &
      9.39e-06 & 8.48e-04 & 4.79e-06 & 1.03e-03 & 6.04e-08 \\ \hline
    \end{tabular}    
  \end{center}
\end{table}

\section{Monte Carlo computation of the distributions of $p$-values}

Although exhaustive computation can derive the exact value of $q_i$'s,   
it is impossible to carry out such a computation 
when the number of realizations is too large. 
For example, 
if we change the parameter of the above Linear Complexity test 
to $m=500$, then the number of 
realizations is increased to be approximately $9.0 \times 10^{16}$,
which is prohibitively large.
For comparison, the number of realizations of the Overlapping Template Matching 
test is approximately $7.2 \times 10^{12}$, 
and the exact $q_i$'s of this test by exhaustive computation 
require 1{,}300 hours on an AMD Ryzen 7 1800X processor 
at 3.6 GHz with 16 GB of memory. 
We thus consider to approximate the values of $q_i$'s by a Monte Carlo simulation. 

Let $\bm{x}_1$, $\ldots$, $\bm{x}_M$ be i.i.d. random samples 
from the sample space $S$ defined by Equation (\ref{eq:samplespace}) in Section 3.
We approximate the value $q_i$ by 
$$
q_i' := \# \left\{ j \mid \mathbb{P}(T(\bm{x}_j) \leq X) \in I_i, 
 j=1, \ldots, M \right\} / M
$$
for $i=0, 1, \ldots, \nu$, where $T$ and $X$ are defined in Section 3.

Figures 1 and 2 show the results of the Monte Carlo simulation 
for computing the chi-squared discrepancy $\delta$ of $\{q'_i\}$ 
from $\{1/(\nu+1)\}$ and 
$u=\max_{i=0, \ldots, \nu}\left| 1- q'_i/p_i \right|$
of the Linear Complexity test with parameter $m=5{,}000$,
respectively:
we use the \texttt{gsl\char`_ran\char`_multinomial} function
in the Gnu Scientific Library (GSL) to generate random samples
$\bm{x}_1$, $\ldots$, $\bm{x}_M$ \cite{galassi2018scientific}.
According to these results, 
we estimate that  $\delta = 4.625 \times 10^{-6}$ and $u=4.442 \times 10^{-3}$, 
and hence $N_{0.25} \approx 507{,}809$ and $N_{0.0001} \approx 5{,}353{,}131$.
Note that the standard deviations of $\delta$ and $u$ in the experiment are
$3.231 \times 10^{-9}$ and $1.837 \times 10^{-6}$, respectively.

\begin{figure}[htpb]
  \includegraphics[width=\textwidth]{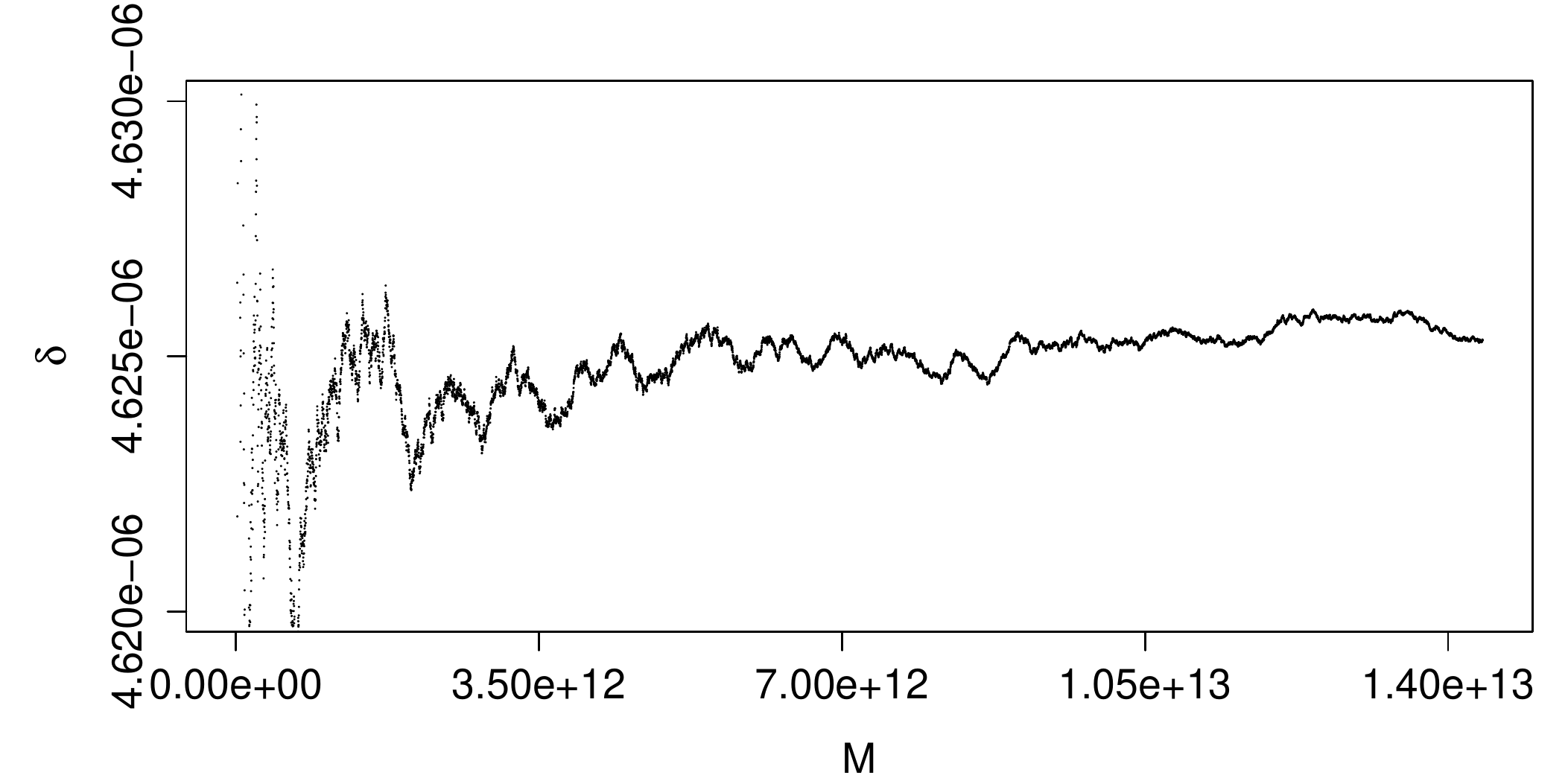}
  \caption{Monte Carlo simulation for $\delta$ of the Linear Complexity test 
  ($m=500$)}

  \includegraphics[width=\textwidth]{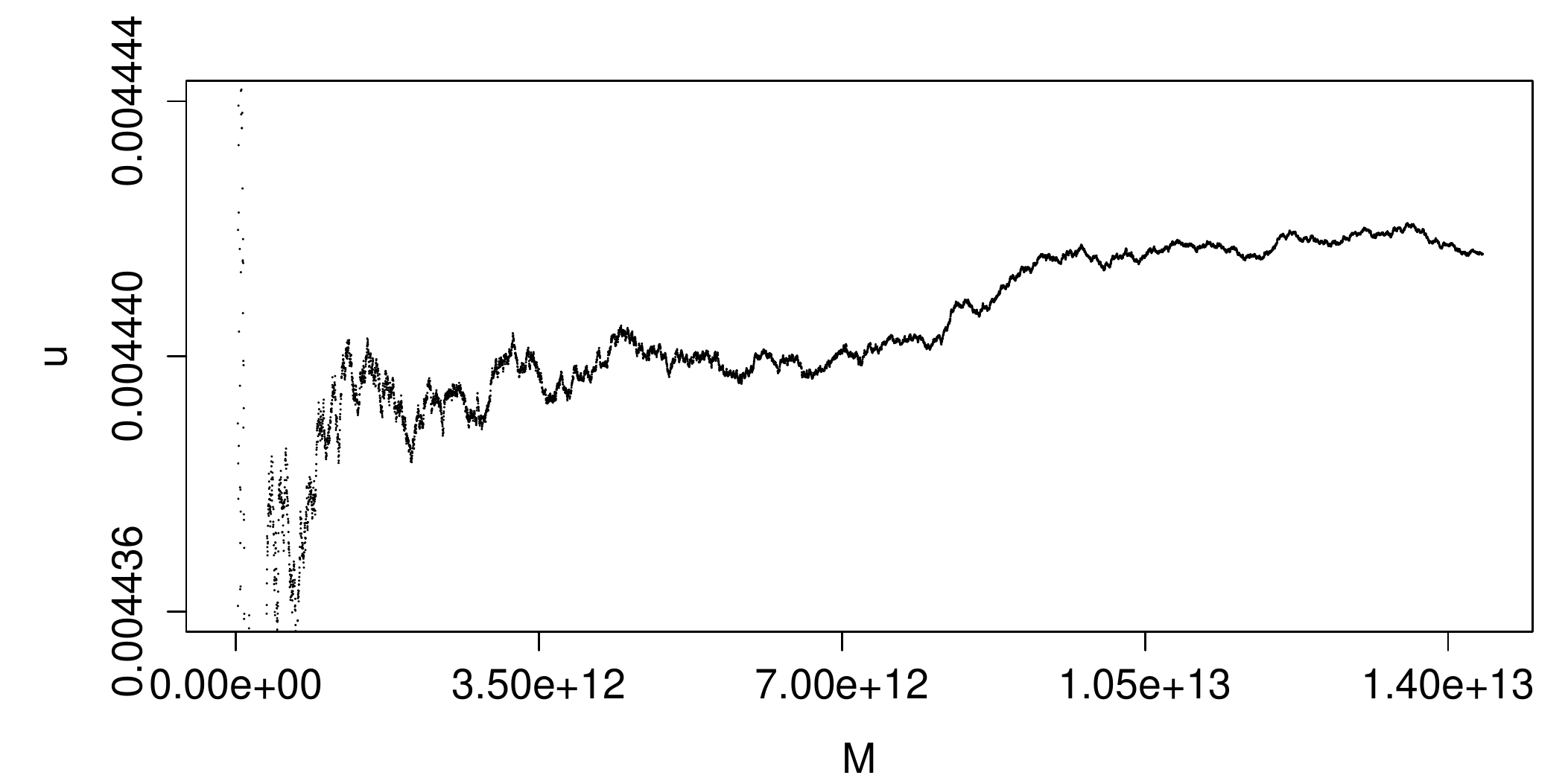}
  \caption{Monte Carlo simulation for 
    $\max_{i=0, \ldots, \nu}\left| 1- q'_i/p_i \right|$ 
    of the Linear Complexity test ($m=500$)}
\end{figure}

Table 6 shows the results of the two-level test with these second sample sizes on SHA-1. 

\begin{table}[htpb]
  \begin{center}
    \caption{$p$-values of the two-level test for the Linear Complexity test ($m=500$)}
    \begin{tabular}{ccccccc}
      \hline 
      PRNG & $N$ & 1st & 2nd & 3rd & 4th & 5th \\ \hline
      SHA-1 & 507{,}000 & 
      2.43e-03 & 7.24e-01 & 2.57e-01 & 8.61e-02 & 5.61e-02  \\ 
      & 5{,}354{,}000 &
      3.73e-14 & 3.19e-09 & 6.73e-08 & 3.25e-09 & 1.49e-11 \\ \hline
    \end{tabular}    
  \end{center}
\end{table}

\paragraph{The Frequency test within a Block.} 
This test divides the given $n$-bit sequence into $n_b = \lfloor n/m \rfloor$ blocks 
of $m$ bits. Let $X_i$ be the number of ones in the $i$-th block for 
$i=1$, $\ldots$, $n_b$. Under $\mathcal{H}_{0,first}$, those $X_1, \ldots, X_{n_b}$ are 
i.i.d. binomial random variables with mean $m/2$ and variance $m/4$. 
The test computes a test statistic 
$$
T(X_1, \ldots, X_{n_b}):=\sum_{i=1}^{n_b} \frac{(X_i-m/2)^2}{m/4} 
= 4m \sum_{i=1}^{n_b} (X_i/m-1/2)^2
$$
which should have approximately the chi-squared distribution  
with $n_b$ degrees of freedom if $m$ is large enough. 

The set of the realizations is $S= \{ 0, 1, \ldots, m \}^{n_b}$
and consequently the total number of realizations is $(m+1)^{n_b}$.  
NIST recommends that $m=128$ and $n \geq 10^6$, 
but exhaustive computation seems impossible, 
so we conduct a Monte Carlo simulation:
in this experiment,we generate random samples
using the \texttt{gsl\char`_ran\char`_binomial} function in
the Gnu Scientific Library.
  
From the results of the Monte Carlo simulation, we approximate that 
$\delta = 2.200 \times 10^{-5}$ and 
$u = 8.990 \times 10^{-2}$ (see Figures 3 and 4.)
The standard deviations of $\delta$ and $u$ are
$1.728\times 10^{-8}$ and $3.775 \times 10^{-6}$, respectively, and 
$N_{0.25} \approx 71{,}802$ and $N_{0.0001} \approx 1{,}160{,}411$. 
Table 7 shows the results of the two-level test on MT and SHA-1. 

\begin{figure}[htpb]
  \includegraphics[width=\textwidth]{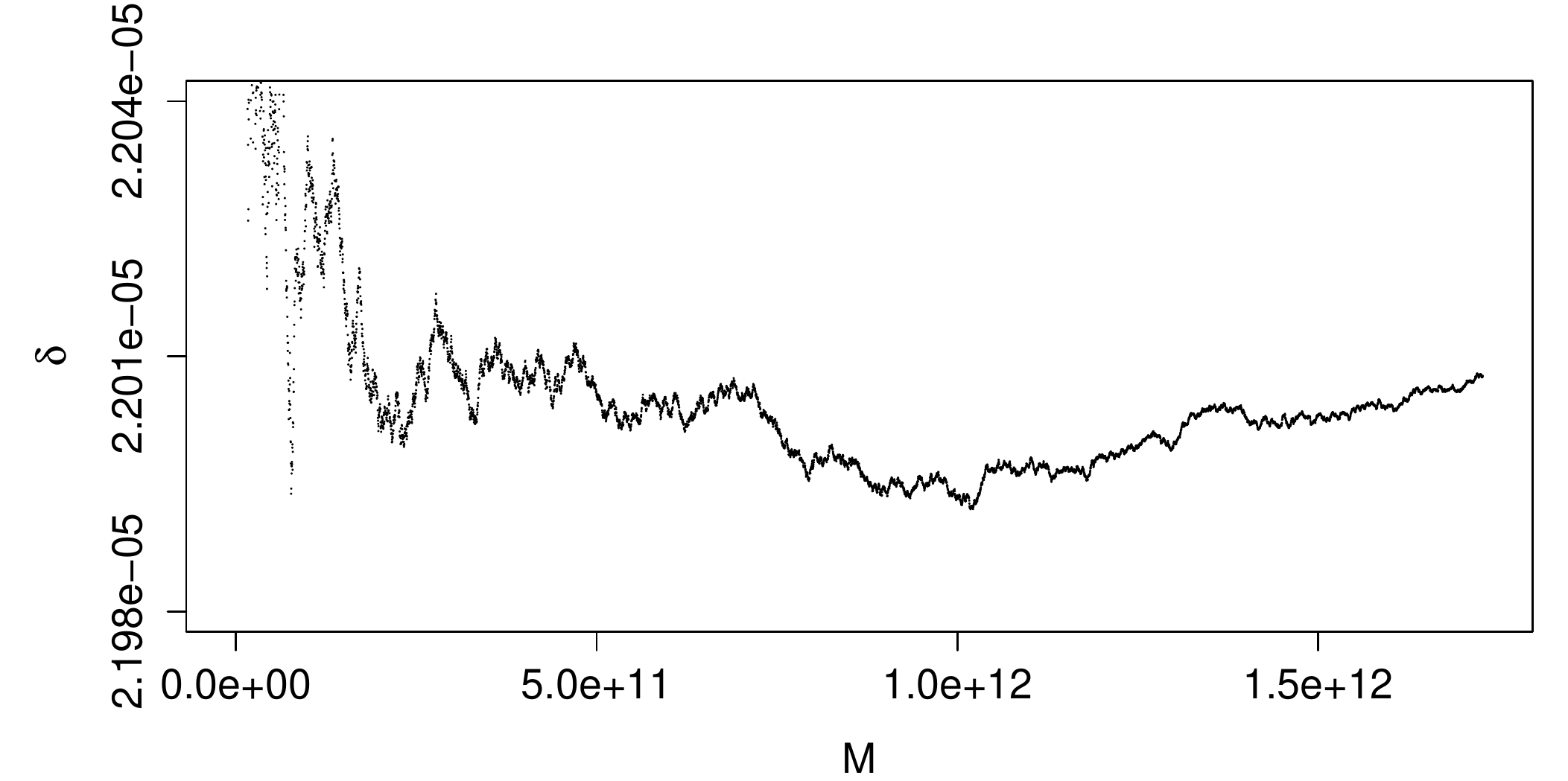}
  \caption{Monte Carlo simulation for $\delta$ of the Frequency test 
  within a Block ($m=128$)}

  \includegraphics[width=\textwidth]{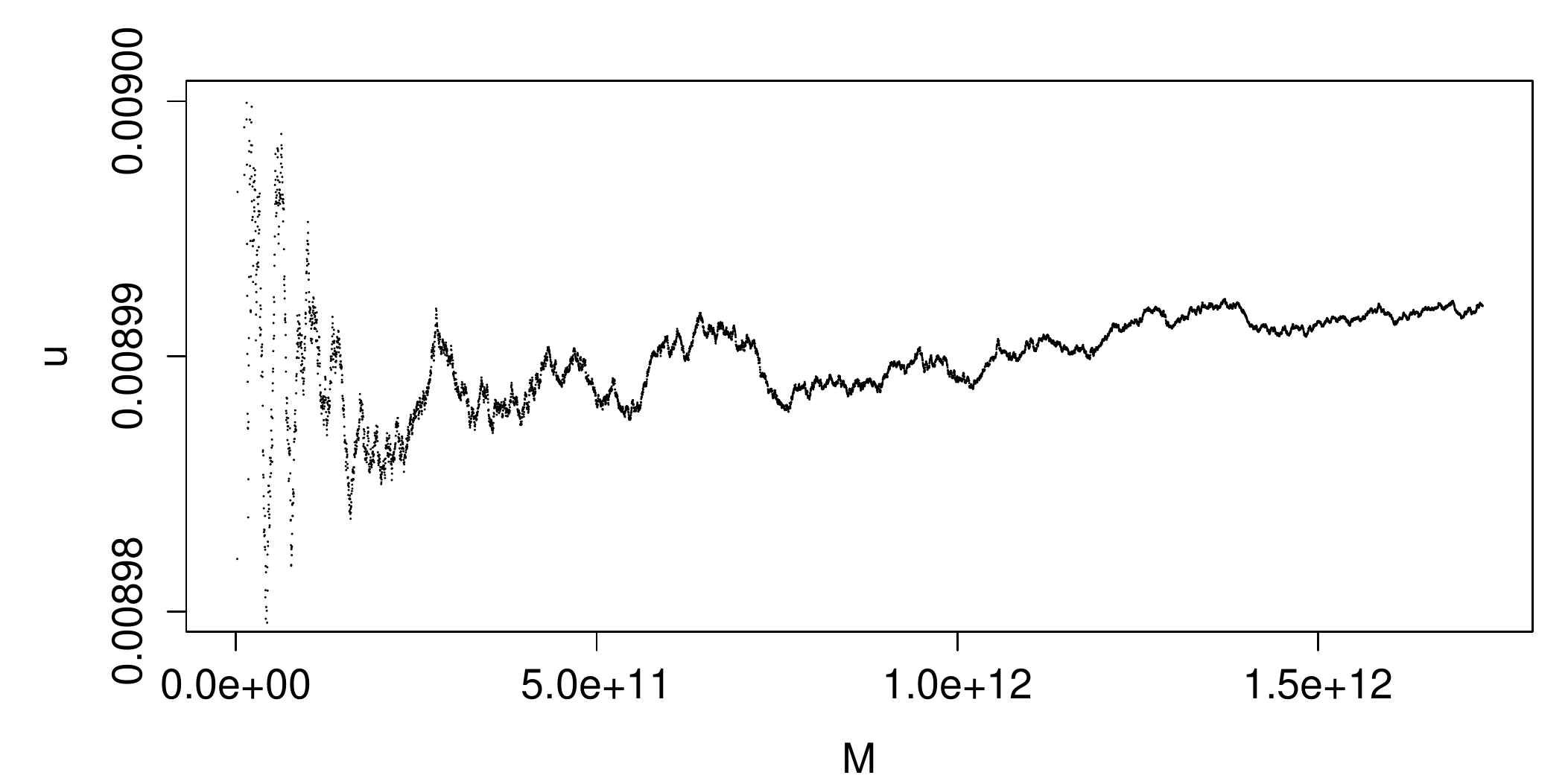}
  \caption{Monte Carlo simulation for 
    $\max_{i=0, 1, \ldots, \nu} \left| 1-q'_i/p_i \right|$ 
    of the Frequency test within a Block ($m=128$)}
\end{figure}

\begin{table}[htpb]
  \begin{center}
    \caption{$p$-values of the two-level test for 
      the Frequency test within a Block  $(m=128)$}
    \begin{tabular}{ccccccc}
      \hline 
      PRNG & $N$ & 1st & 2nd & 3rd & 4th & 5th \\ \hline
      MT & 71{,}800 & 
      3.06e-02 & 5.21e-04 & 5.64e-01 & 6.85e-01 & 4.68e-01 \\
      & 1{,}161{,}000 &
      8.20e-09 & 3.65e-06 & 6.44e-06 & 9.28e-06 & 1.80e-03 \\ \hline
      SHA-1 & 718{,}00 & 
      9.60e-01 & 7.21e-01 & 3.74e-01 & 6.51e-01 & 1.72e-01  \\ 
      & 1{,}161{,}000 &
      2.35e-04 & 1.06e-05 & 1.48e-03 & 9.52e-03 & 1.69e-07 \\ \hline
    \end{tabular}    
  \end{center}
\end{table}

These experimental results indicate that the Monte Carlo simulation
gives good approximations of $\delta$, $q_i$'s,
and the risky/safe sample sizes. 

\paragraph{The Discrete Fourier Transform (DFT) test.} 
The purpose of this test is to detect periodic features in the tested 
sequence that would indicate a deviation from the assumption of randomness.

The first sample size $n$ of the DFT test must be even. 
The discrete Fourier coefficients for the random bit variables $B_k$ 
are defined by
$$
F_i = \sum_{k=0}^{n-1} (2B_k-1) \exp(-2\pi \sqrt{-1} k i/ n)
$$
for $i=0,1, \ldots, n/2-1$. 
Let $O_h$ denote the observed number of $|F_i|$'s 
that are smaller than a specified threshold $h$.

NIST SP800-22 falsely assumes that $|F_i|$'s are mutually independent.
Under this assumption and when $h=\sqrt{2.995732274n}$,
$O_h$ is approximately a binomial random variable 
with $n/2$ trials and probability $0.95$, 
and $O_h$ has approximately the normal distribution 
with mean $\mu=0.95n/2$ and variance $\sigma_0^2 = 0.05 \cdot 0.95 n/2$. 
Hence, NIST SP800-22 defines the approximated $p$-value of the DFT test 
for a realization $o_h$ of $O_h$ by
$$
\mathbb{P} (|o_h-\mu|/\sigma_0 < |X|)
= 2(1-\mathrm{\Phi}(|o_h-\mu|/\sigma_0)),
$$
where $X$ is a random variable that conforms to the standard normal distribution $N(0,1)$ 
and $\mathrm{\Phi}$ is the cumulative distribution function 
of the standard normal distribution. 
Since $o_h$ varies from $0$ to $n/2$, we have 
$$
\displaystyle q_i= \sum_{0 \leq j \leq n/2, \mathbb{P}(|j-\mu|/\sigma_0 < |X|) \in I_i} 
\binom{n/2}{j} \times 0.95^j \times 0.05^{n/2-j}.
$$

For the first sample size $n=10^6$, we have 
$N_{0.25} = 18{,}690$, and $N_{0.0001} = 210{,}628$. 

Actually, the approximation of $O_h$ by the normal distribution 
$N(\mu, \sigma_0^2)$ is inaccurate, 
and there are many reports on defects of the DFT test. 
Indeed, NIST corrected the approximation of the variance to a better 
value $\sigma_1^2 = 0.05 \cdot 0.95 n/4$ 
proposed by Kim et al. \cite{journals/iacr/KimUH04} from $\sigma_0^2$. 
Later, Pareschi et al. \cite{6135498} found a further good approximation
value $\sigma_2^2=0.05 \cdot 0.95n/3.8$. 
According to \cite{HARAMOTO201966}, we use $\sigma_2^2$ for variance
in the following experiments.

Next, we consider the following Monte Carlo simulation. 
Using MT, we generate $M$ $n$-bit sequences and 
yield $M$ $p$-values by the DFT test. We count the number of those $p$-values 
that fall in the interval $I_i$, $i=0, 1, \ldots, \nu$.
Then we use this empirical distribution instead of the actual distribution.

Figures 5 and 6 show the results of the Monte Carlo simulations. 
We approximate $\delta$ by $5.438 \times 10^{-4}$  and $u$ by $3.679 \times 10^{-2}$ 
using the Monte Carlo simulation with $M=1.7 \times 10^{12}$ sequences:
the standard deviations are $7.419 \times 10^{-7}$ and
  $6.339 \times 10^{-5}$. 
The results are $N_{0.25}=3{,}785$ and $N_{0.0001}=46{,}084$.
Table 8 shows the $p$-values of the two-level test 
with the second sample sizes of $3{,}700$, $18{,}600$, $46{,}100$ and $211{,}000$.
In the experiments, we test on SHA-1 and WELL \cite{Panneton:2006:ILG:1132973.1132974}:
we avoid to test on MT because it is used
for generating random sequences for approximating $\delta$ and $u$.
These results indicate that the Monte Carlo simulation gives 
an approximation value of $\delta$ with enough accuracy in practical use. 

\begin{figure}[htpb]
  \includegraphics[width=\textwidth]{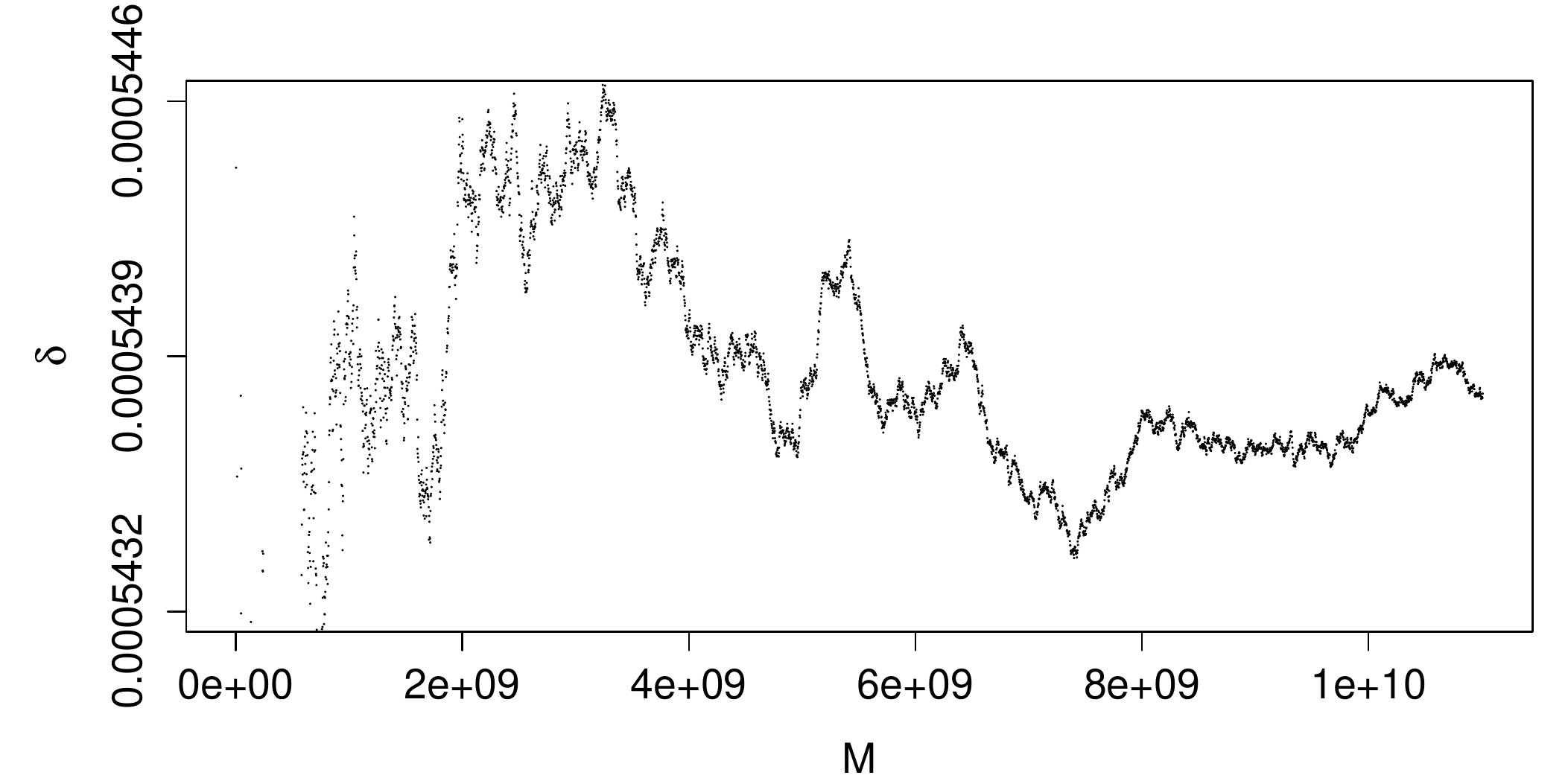}
  \caption{Monte Carlo simulation for $\delta$ of the DFT test}

  \includegraphics[width=\textwidth]{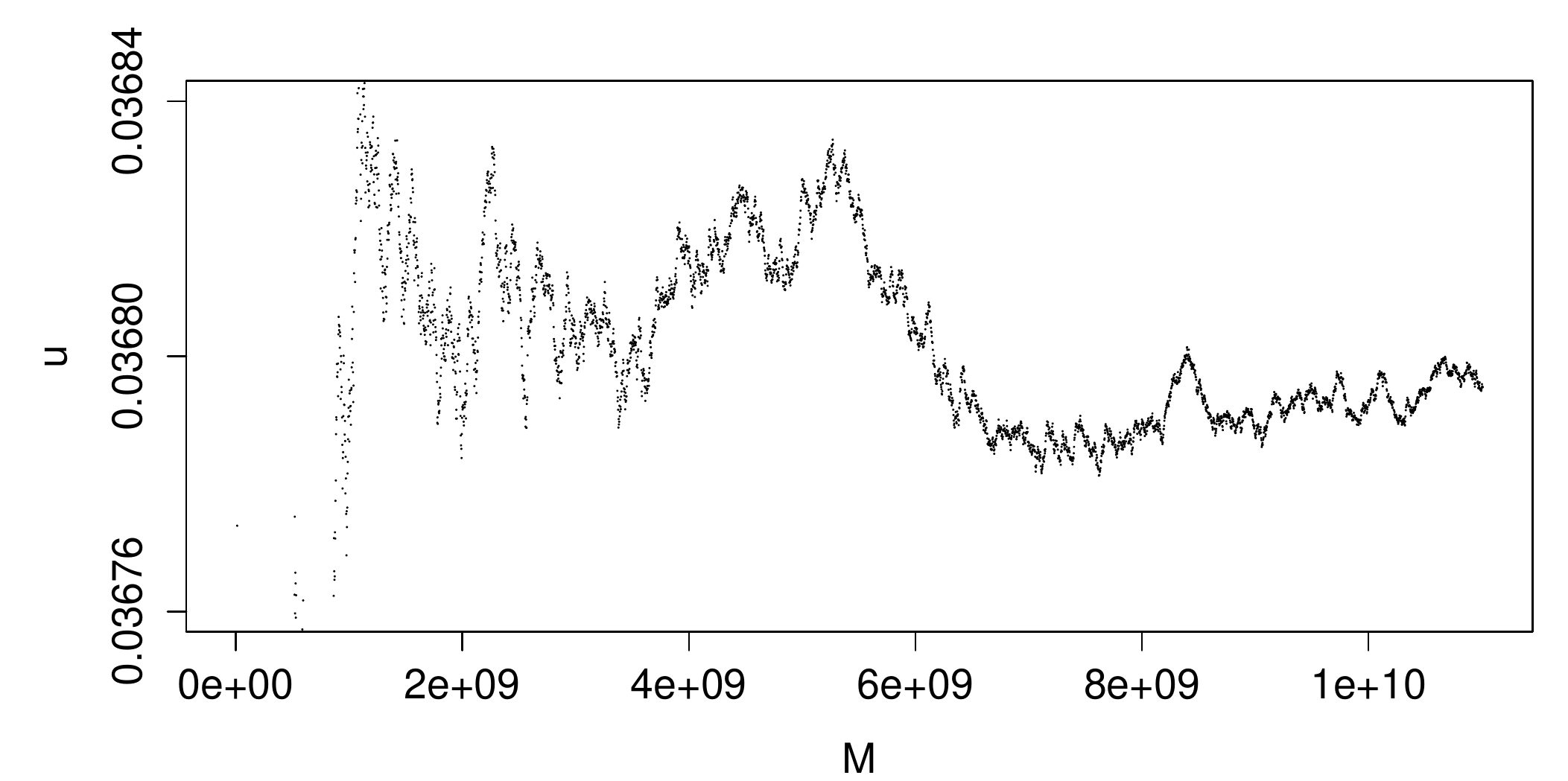}
  \caption{Monte Carlo simulation for 
    $\max_{i=0,1,\ldots,\nu} \left| 1-q'_i/p_i \right|$
   of the DFT test}
\end{figure}

\begin{table}[htpb]
  \begin{center}
    \caption{$p$-values of the two-level test for the DFT test}
    \begin{tabular}{ccccccc}
      \hline 
      PRNG & $N$ & 1st & 2nd & 3rd & 4th & 5th \\ \hline
      WELL & 3{,}700 & 
      6.20e-01 & 1.58e-01 & 4.23e-03 & 2.28e-02 & 3.59e-01  \\ 
      & 18{,}600 &
      8.33e-02 & 4.22e-01 & 1.55e-01 & 3.20e-01 & 5.47e-01  \\ 
      & 46{,}100 &
      4.51e-03 & 6.00e-02 & 1.29e-02 & 2.06e-02 & 1.43e-01  \\ 
      & 211{,}000 &
      3.60e-22 & 9.00e-16 & 2.12e-18 & 9.13e-20 & 2.37e-14 \\ \hline
      SHA-1 & 3{,}700 &
      8.75e-01 & 4.02e-01 & 5.07e-01 & 1.80e-01 & 3.77e-01  \\ 
      & 18{,}600 & 
      1.06e-01 & 1.85e-01 & 6.71e-02 & 3.98e-08 & 4.67e-03  \\ 
      & 46{,}100 &
      5.46e-03 & 1.24e-03 & 2.07e-03 & 4.29e-06 & 9.41e-07  \\ 
      & 211{,}000 &
      6.36e-22 & 4.59e-22 & 2.18e-20 & 4.39e-25 & 3.94e-25 \\ \hline
    \end{tabular}    
  \end{center}
\end{table}

\section{Improving two-level testing}

As explained in Section 2, the two-level test under the null hypothesis
$$
\mathcal{H}_{0, second} : p_0 = 1/(\nu+1),  \ldots, p_{\nu} = 1/(\nu+1), 
$$
tends to give erroneous rejections when the second sample size is too large. 
Consequently the second sample size must be limited 
and we recommend $N_{0.25}$ as an upper limit.
This restriction is necessary but inconvenient.  

However, if we use a null hypothesis 
$$
\mathcal{H}'_{0,second} : p_0 = q_0, \ldots ,p_{\nu} = q_{\nu}
$$
where $\{ q_i \}$ is the actual distribution of approximated $p$-values,
then the chi-squared discrepancy $\delta$ becomes $0$. 
This implies that we can increase the value of upper limit 
of the two-level test. 
  
For example, in Section 3, we derived $\{q_i\}$ of 
the following four tests: the Longest-Run-of-Ones in a Block, 
Overlapping Template Matching test, 
Linear Complexity test with parameter $m=5{,}000$, and the Random Excursions test.
We then apply the two-level test under $\mathcal{H}'_{0,second}$ to MT and SHA-1.  

Table 9 shows the $p$-values of several empirical results.
The first sample size is $n=10^6$, and the second sample size $N$ is shown 
in the third column of Table 9. 
These $N$'s are approximately $N_{0.0001}$ or $2N_{0.0001}$:
recall that $N_{0.0001}$ is the risky sample size under $\mathcal{H}_{0,second}$. 

In the experiments, we observe that the $p$-values are moderate,
thus these results are as we expected.

\begin{table}[htpb]
  \begin{center}
    \caption{$p$-values of the two-level test under $\mathcal{H}'_{0,second}$}
    \begin{tabular}{ccccccccc}
      \hline 
      Test & PRNG & $N$ & 1st & 2nd & 3rd & 4th & 5th \\ \hline
      Longest & MT & 500{,}000 &
      9.36e-02 & 4.16e-01 & 9.48e-01 & 6.01e-01 & 5.84e-01 \\ 
      & SHA-1  & 500{,}000 &
      5.91e-01 & 1.31e-01 & 2.96e-01 & 3.80e-01 & 5.19e-01 \\ \hline 
      Overlap & MT & 27{,}033{,}000 &
      3.24e-02 & 2.58e-02 & 4.66e-03 & 5.83e-01 & 1.27e-01 \\
      & SHA-1 & 27{,}033{,}000 & 
      4.36e-01 & 4.29e-01 & 7.57e-01 & 2.87e-01 & 3.00e-04 \\ \hline
      Linear & SHA-1 & 100{,}000 &
      7.42e-01 & 1.40e-01 & 4.00e-01 & 7.94e-01 & 2.51e-01 \\ \hline
    \end{tabular}
  \end{center}
\end{table}

Table 10 shows the results of the two-level test for the Random Excursions test 
when $n=10^6$ and $N=2 \times 10^6$ under $\mathcal{H}'_{0,second}$. 
We show the $p$-value less than the significance level $0.0001$ 
and test parameter $x$: the symbol ``-'' indicates that no rejection occurred. 

The results indicate that, under $\mathcal{H}'_{0,second}$, the two-level test 
seems sufficiently accurate even if we take a larger sample size.

\begin{table}[htpb]
  \begin{center}
    \caption{Results of the two-level test for the 
      Random Excursions test under $\mathcal{H}'_{0,second}$}
    \begin{tabular}{ccccccc}
      \hline 
      PRNG & $N$ & 1st & 2nd & 3rd & 4th & 5th \\ \hline
      MT & 2{,}000{,}000 & - & - & - & - & - \\ 
      SHA-1 & 2{,}000{,}000 & 3.77e-05 ($x=-3$) & - & - & - & 1.27e-05 ($x=1$) \\ \hline
    \end{tabular}
  \end{center}
\end{table}

In Section 4, we approximate distributions of approximated $p$-values, say $\{q'_i\}$, 
using a Monte Carlo simulation. We consider to replace the null hypothesis 
$\mathcal{H}_{0,second}$ with
$$
\mathcal{H}''_{0,second} : p_0 = q'_0, \ldots, p_{\nu} = q'_{\nu} 
$$
to make upper limit of second sample sizes more flexible. 

We apply the two-level test under $\mathcal{H}''_{0,second}$ to MT and SHA-1. 
The values of $q'_i$'s are presented in Table 11: these values are
calculated by the Monte Carlo simulation described in Section 4. 
The first sample sizes is $n=10^6$ and the second sample size $N$ is 
indicated in the third column of Table 12. Similar to the previous experiments, 
we took $N_{0.0001}$ or $2N_{0.0001}$ as the second sample sizes. 

Table 12 shows the $p$-values of the two-level test. From these results, 
the null hypothesis $\mathcal{H}''_{0,second}$ also allows us to take larger sample sizes.

\begin{table}[htpb]
  \begin{center}
    \caption{Values of $q'_i$'s by Monte Carlo simulation}
    \begin{tabular}{cccc}
      \hline
      Test & Linear ($m=500$) & Block Freq. & DFT \\ \hline
      $q'_0$ & 0.09985 & 0.09912 & 0.1012 \\
      $q'_1$ & 0.09956 & 0.09993 & 0.0987 \\
      $q'_2$ & 0.09983 & 0.10029 & 0.1023 \\
      $q'_3$ & 0.09999 & 0.10028 & 0.0964 \\
      $q'_4$ & 0.10016 & 0.10042 & 0.1015 \\
      $q'_5$ & 0.10026 & 0.10039 & 0.1012 \\
      $q'_6$ & 0.10025 & 0.10027 & 0.0963 \\
      $q'_7$ & 0.10017 & 0.10024 & 0.1014 \\
      $q'_8$ & 0.10009 & 0.09996 & 0.0979 \\
      $q'_9$ & 0.09984 & 0.09910 & 0.1031 \\ \hline
    \end{tabular}    
  \end{center}
\end{table}

\begin{table}[htpb]
  \begin{center}
    \caption{$p$-values of the two-level tests under $\mathcal{H}''_{0,second}$} 
    \begin{tabular}{ccccccccc}
      \hline 
      Test & PRNG & $N$ & 1st & 2nd & 3rd & 4th & 5th \\ \hline
      Linear & SHA-1 & 5{,}354{,}000 & 
      1.68e-04 & 3.09e-02 & 2.41e-01 & 5.07e-02 & 5.11e-03 \\ \hline 
      Block Freq. & MT & 2{,}000{,}000 &
      3.54e-02 & 2.50e-01 & 9.63e-01 & 4.46e-01 & 7.60e-01 \\ 
      & SHA-1  & 2{,}000{,}000 &
      1.68e-01 & 3.68e-01 & 9.02e-01 & 7.97e-01 & 6.90e-01 \\ \hline
      DFT  & WELL & 100{,}000 &
      8.97e-01 & 1.03e-01 & 9.99e-01 & 1.87e-01 & 8.71e-01 \\ 
      & SHA-1 & 100{,}000 &
      8.01e-01 & 9.51e-01 & 8.82e-01 & 9.75e-01 & 5.92e-01 \\ \hline
    \end{tabular}
  \end{center}
\end{table}

\section{Conclusions}
In this paper, we study practical upper limits of the sample sizes 
of the two-level test. 
With our previous work \cite{10.1007/978-3-319-91436-7_15}, 
we derive upper limits of the two-level test for nine one-level tests 
in NIST SP800-22.  
The upper limits proposed in this paper reveals that 
NIST's recommendation of the second sample size,
the order of the inverse of the significance level, might be misleading:
appropriate order of the second sample size strongly depends on 
the one-level test and the first sample size. 

A future work is to derive upper limits or the distributions of 
approximated $p$-values of the remaining six one-level tests. 
Monte Carlo simulation would be suitable for those six tests,  
but requires a large amount of computation time.
To conduct a sufficiently accurate Monte Carlo simulation, 
we need to compute the $p$-values of one-level tests more efficiently 
than the current NIST SP800-22. 

\begin{acknowledgements}
  The author is indebted to Professor Makoto Matsumoto for constant help
  and encouragements.
  The author is thankful to the anonymous referees
  for many valuable comments.

  This study was carried out under the ISM Cooperative Research Program 
  (2018-ISMCRP-10 and 2019-ISMCRP-05).
\end{acknowledgements}

%
\section*{Conflict of interest}

The authors declare that they have no conflict of interest.

\bibliographystyle{spmpsci}      
\bibliography{haramoto01}   

\end{document}